# Acoustic source depth estimation method based on a single hydrophone in Arctic underwater

Jinbao Weng[1,2,1], Yubo Qi[3], Yanming Yang[1,2], Hongtao Wen[1,2], Hongtao Zhou[1,2], Benqing Chen[1,2], Dewei Xu[1,2], Ruichao Xue[1,2], Caigao Zeng[1,2]

1. Laboratory of Ocean acoustics and Remote Sensing, Third Institute of Oceanography, Ministry of Natural Resources, Xiamen, Fujian 361005, China
2. Fujian Provincial Key Laboratory of Marine Physical and Geological Processes, Xiamen, Fujian 361005, China
3. State key laboratory of acoustics, Institute of Acoustics, Chinese Academy of Sciences, Beijing 100190, China

## Abstract

Based on the normal mode and ray theory, this article discusses the characteristics of surface sound source and reception at the surface layer, and explores depth estimation methods based on normal modes and rays, and proposes a depth estimation method based on the upper limit of modal frequency. Data verification is conducted to discuss the applicability and limitations of different methods. For the surface refracted normal mode waveguide, modes can be separated through warping transformation. Based on the characteristics of normal mode amplitude variation with frequency and number, the sound source depth can be estimated by matching amplitude information. Based on the spatial variation characteristics of eigenfunctions with frequency, a sound source depth estimation method matching the cutoff frequency of normal modes is proposed. For the deep Arctic sea, the sound ray arrival structure at the receiving end is obtained through the analysis of deep inversion sound ray trajectories, and the sound source depth can be estimated by matching the time difference of ray arrivals. Experimental data is used to verify the sound field patterns and the effectiveness of the sound source depth estimation method.

---

[1] Email: wengjinbao@tio.org.cn

# 1 Introduction

Due to its unique marine environment, the Arctic Ocean exhibits a sound speed profile distinct from that of conventional sea areas, leading to sound field characteristics that differ from those of conventional shallow or deep seas. Since the 1960s, researchers have conducted experimental studies on sound propagation and source localization in the Arctic Ocean. In the 1980s, T C Yang and others utilized a large-aperture vertical array to separate sound field modes and achieved source distance and depth estimation through the matched field method.

In recent decades, the Arctic marine environment has undergone significant changes, including the rise in sea temperature in the Arctic Ocean and the intrusion of warm Pacific water into the Arctic Ocean, resulting in the formation of a dual-channel sound speed profile in the Canadian Basin and the Chukchi Plateau. In such a rapidly changing Arctic environment, studying the corresponding sound propagation phenomena and sound source localization methods remains innovative.

Currently, there are numerous methods for estimating the depth of a sound source based on a single hydrophone. These include methods based on normal mode theory, ray theory, and others. Interference structures and time difference of arrival (TDOA) are also utilized. Commonly, the amplitude variation of normal modes is used to estimate depth in shallow waters, while the TDOA of rays is used to estimate depth in deep waters, or it manifests as interference structures in the frequency domain.

Over the past decade, numerous studies have been conducted on the separation of normal modes using warping transformations, and further research has been carried out on single hydrophone ocean environment inversion and sound source localization based on this foundation. However, current research mainly focuses on seabed reflection-type normal modes in shallow waters, and there is relatively little research on warping to achieve modal separation for refraction-type normal modes formed in the surface layer of the deep Arctic sea, which could then be used for environmental inversion and sound source localization.

Currently, there is limited discussion on the depth estimation of underwater sound sources in the Arctic. This article attempts to explore the depth estimation of sound sources within the surface layer of the deep Arctic, analyzing the characteristics of the sound field and acoustic signals. It discusses depth estimation methods based on normal modes and rays, respectively, and validates them through experimental data.

## 2 Theoretical analysis and depth estimation methods

### A. Characteristics of acoustic signals received at different depths in the deep Arctic waters

Under Arctic deep-sea conditions, the received acoustic signals below the surface layer depth consist of two parts: the ray multipath arrival component and the normal mode multimodal arrival component. Among them, the ray multipath arrival component has a faster equivalent sound speed due to its experience of water bodies over a large depth range, thus arriving earlier; the normal mode multimodal arrival component, on the other hand, arrives later because its main energy is concentrated in the surface seawater where the sound speed is lower. Therefore, on the time-domain waveform, one can observe the non-dispersive ray multipath arrival component and the dispersive normal mode multimodal arrival component arriving in succession.

Under Arctic deep-sea conditions, the received acoustic signals at mid-depths and seabed depths are primarily characterized by ray multipath arrivals. Since the equivalent sound velocities of water bodies over the large depth range experienced by sound rays at mid-depths and seabed depths are essentially consistent, the fewer the number of large depth inversions, the shorter the sound ray path length, resulting in the earliest arrival. Moreover, subsequent arriving sound rays need to undergo seabed reflection, so the earliest arriving sound ray has the shortest path length and the highest energy, and the intensity of the acoustic signal gradually decreases with time.

### B. Depth estimation method based on normal mode amplitude

According to the normal mode theory, due to the characteristic of vertical distribution of eigenfunctions fluctuating with depth, the amplitude of normal modes excited by different source depths is also inconsistent. The amplitude of each normal mode varies with source depth and also with source frequency. Therefore, the depth can be estimated by matching the amplitude variation with frequency.

## C. Depth estimation method based on the upper frequency limit of normal modes

Under Arctic conditions, as the vertical distribution of the eigenfunction for each normal mode gradually decreases with increasing frequency, when the depth of the sound source exceeds the vertical distribution range of the eigenfunction, the sound source cannot excite that normal mode. This phenomenon results in a frequency upper limit for the mode at a given sound source depth. By calculating the curve of the modal frequency upper limit as a function of the sound source frequency under known environmental conditions, the sound source depth can also be estimated by matching the upper limit.

## D. Depth estimation method based on ray theory

Based on the time difference of arrival (TDOA), which refers to the time difference between the four rays originating from the sound source, reflecting off the sea surface, and being received and reflected again off the sea surface, the sound source depth and distance can be estimated using the TDOA, provided that the receiving depth is known.

## 3 Introduction to Arctic experiment

## A. Experimental setup and marine environmental information

In the Chukchi Plateau sea area, an experiment on sound source depth estimation was conducted using pulsed sound sources. The locations of the five sound source deployment stations and the receiving station are shown in the figure, with propagation distances ranging from 105km to 1066km. The sound source depths included two types: 100m and 300m. Acoustic signals were received using acoustic buoys, with a receiving depth of 342m, employing a self-contained underwater acoustic receiving device with a single hydrophone. The seabed topography along the sound propagation path was obtained from the ETOP database. As can be seen from the figure, the receiving buoy is located at a water depth of approximately 2000m. The topography changes significantly along the path from the five sound source locations to the receiving location. Especially at station S22 where the water depth is only a few hundred meters, transitioning from a shallow sea environment to a deep sea environment. Sound source stations S11, ICE05, ICE06, and ICE08 are all located in deep sea areas. Satellite remote sensing data was used to obtain information on sea ice distribution. As can be seen from the figure, there is no sea ice coverage along the path from stations S11 and S22 to the receiving station, while there is extensive sea ice coverage along the path from stations ICE05, ICE06, and ICE08 to the receiving station. Additionally, during the experiment, sound velocity profiles were measured at multiple stations. Stations at shorter distances mainly exhibited dual-channel sound velocity profiles, while stations at longer distances mainly exhibited half-channel sound velocity profiles.

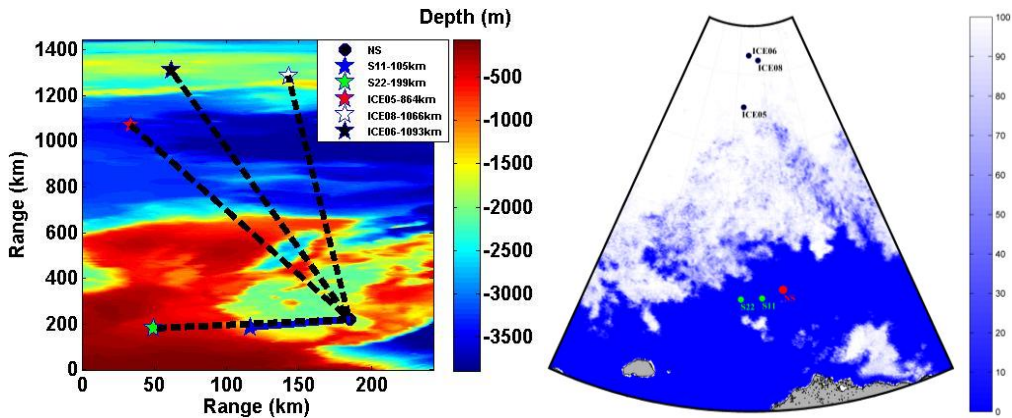

Figure 1 Submarine topography and sea ice coverage

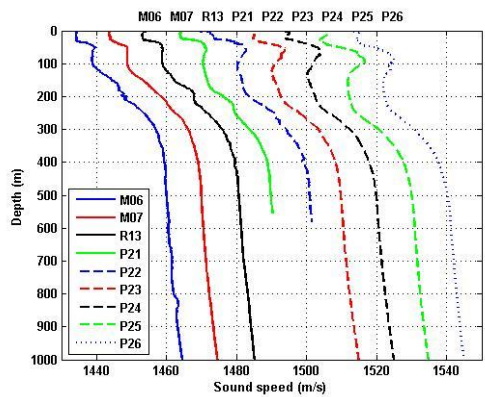

Figure 2 Seawater sound velocity profile

## B. Sound field simulation under experimental conditions

Acoustic field simulations were conducted for acoustic signals at four different distances. As can be seen from the figure, for the first acoustic propagation path, when there is a large depth of the sea along the propagation path, the acoustic field simultaneously exhibits a surface normal mode acoustic field and a deep-sea large-depth-reversed ray acoustic field. Therefore, the received acoustic signals in the surface layer should observe both ray arrival and normal mode arrival structures. For the second acoustic propagation path, the seabed topography changes complexly, transitioning from shallow water around 400m to deep water around 2000m. At this time, the deep-sea large-depth-reversed sound rays are blocked by the seabed and cannot propagate. Consequently, the surface acoustic field is dominated by normal mode arrivals, and the received acoustic signals are also primarily characterized by multiple normal mode arrivals. Therefore, the long-range propagation of deep-sea large-depth-reversed sound rays is limited by seabed topography, while the long-range propagation of surface normal modes requires less seabed topography and is easier to achieve, indicating that positioning methods based on the surface normal mode acoustic field are less constrained by seabed topography variations.

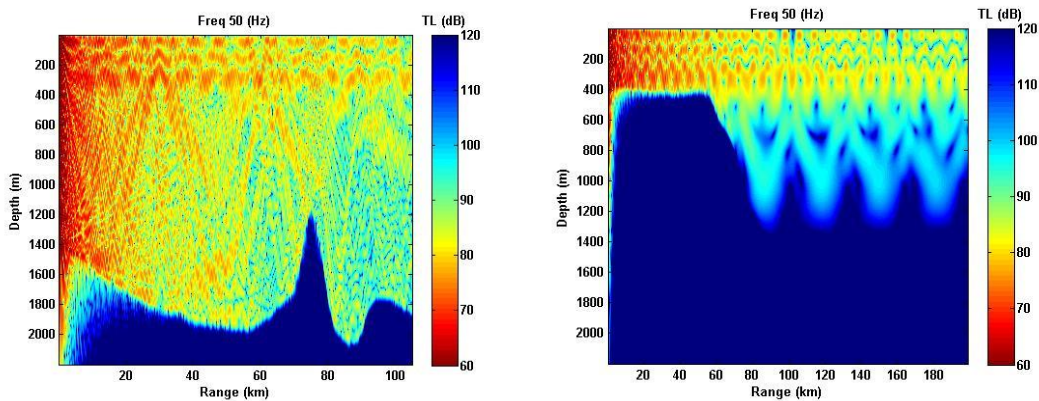

Figure 3 Acoustic field simulation results

## C. Receiving acoustic signal time-domain waveform and time-frequency analysis

For the received acoustic signals, time-domain waveform and time-frequency analysis are conducted to analyze the multi-path arrival information and multi-mode arrival structure. The time-domain waveform and time-frequency analysis of the received acoustic signals at a sound source depth of 300m are shown in the figure below. It can be seen from the figure that the received acoustic signals at a distance of 105km contain multi-path structures of sound rays without dispersion that arrive first, as well as multi-mode arrival structures with dispersion that arrive later, which is consistent with the conclusions of the two-dimensional acoustic field simulation analysis. It can also be seen from the figure that the received acoustic signals at a distance of 199km only contain multi-mode arrival structures with dispersion, which is also consistent with the conclusions of the two-dimensional acoustic field simulation analysis. Therefore, for deep-water propagation, positioning can be carried out using the multi-path sound rays and multi-mode arrivals inverted at great depths; for complex underground conditions, positioning can only be carried out using the multi-mode propagation in the surface layer.

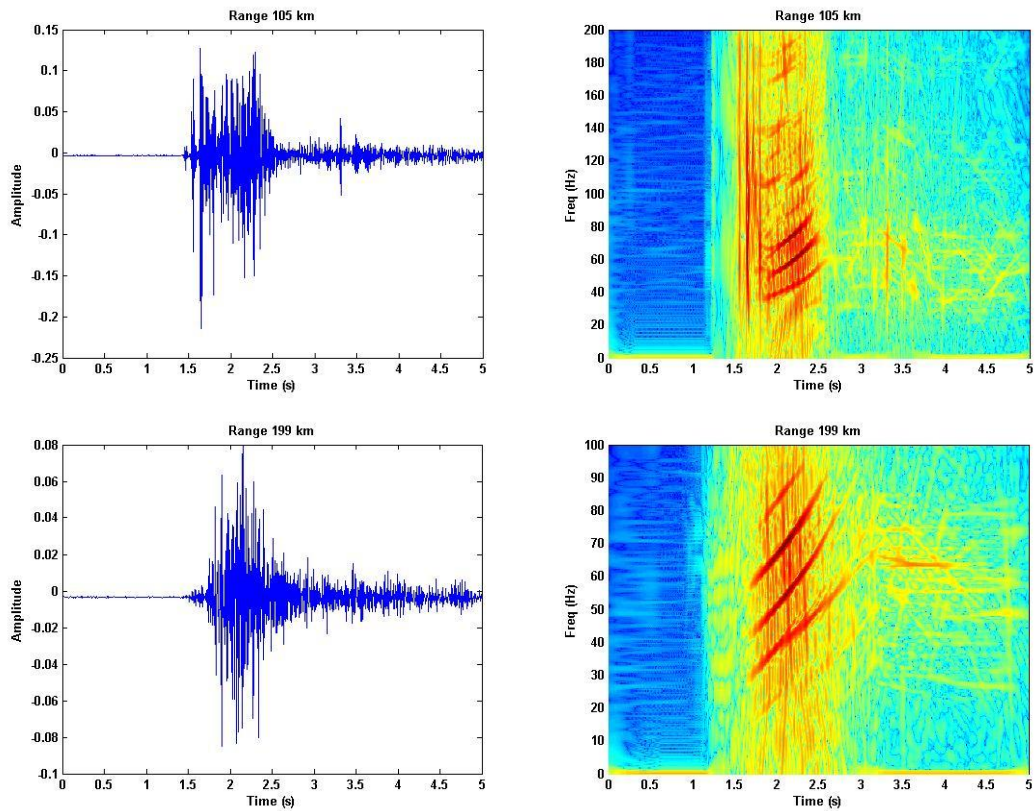

Figure 4 Time-domain waveform and time-frequency analysis (sound source depth of 300m, propagation distances of 105km and 199km)

When the sound source depth is 100m, the time-domain waveform and time-frequency analysis of acoustic signals at two propagation distances are presented in the figure below. For the case of a propagation distance of 105km, multipath ray arrivals and multimodal arrivals can still be observed in the received acoustic signals. Unlike the case of a sound source depth of 300m, the distribution of normal mode energy varies significantly with the mode number and frequency, mainly due to changes in the amplitude of excited modes caused by variations in sound source depth, which is consistent with theoretical analysis. For the case of a propagation distance of 199km, compared to the case of a sound source depth of 300m, the distribution of normal mode energy with mode number and frequency changes significantly, with almost no energy arriving for the third and sixth normal modes. According to the vertical distribution of the eigenfunctions, the sound source depth of 100m may be located exactly at the node positions of the third and sixth normal modes, thus preventing the excitation of these two normal modes. By comparing the acoustic signals at different sound source depths, it can be seen that the sound source depth can be estimated by matching the variation pattern of normal mode energy with frequency for different mode numbers.

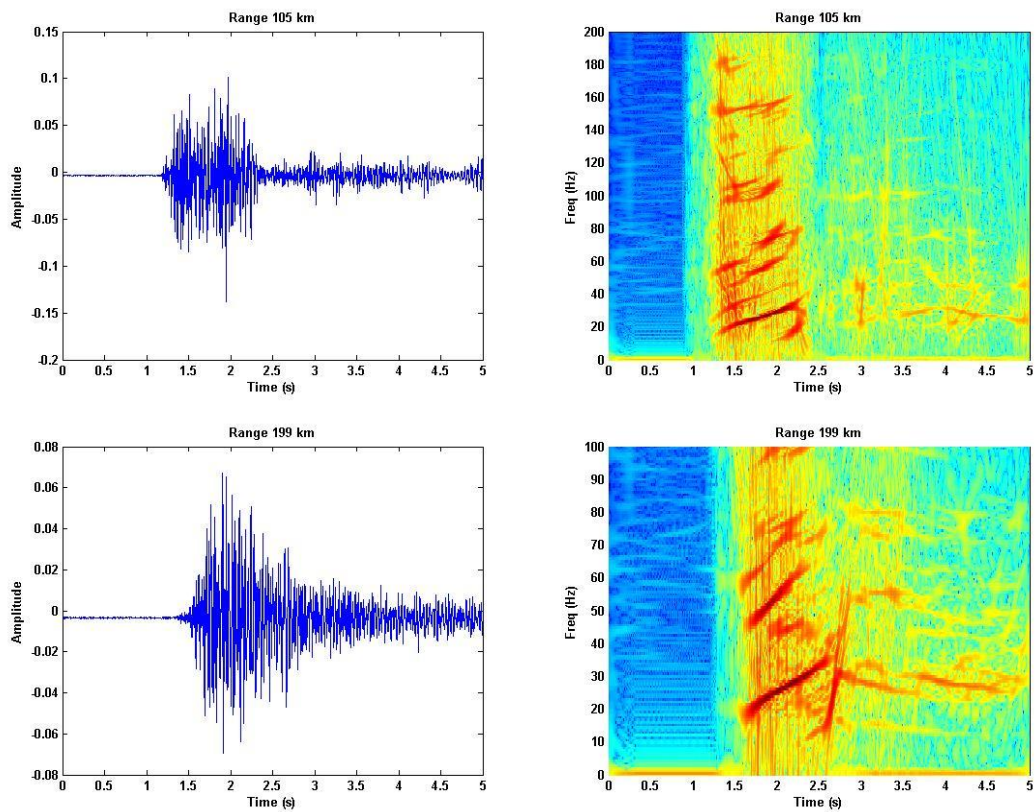

Figure 5 Time-domain waveform and time-frequency analysis (sound source depth of 100m, propagation distances of 105km and 199km)

## D. Modal separation and dispersion curve extraction of received acoustic signals

To estimate the depth information of the sound source, it is necessary to carry out modal separation work to obtain the amplitude information of each normal mode. For the received acoustic signals in this experiment, the warping variation of refractive normal modes is utilized to transform each normal mode into a single-frequency signal, as shown in the figure below. It can be seen from the figure that each normal mode has been transformed into a single-frequency signal, while the multi-path arrival structure does not transform into a single-frequency signal due to the absence of dispersion effect. Subsequently, bandpass filtering and inverse warping variation can be utilized to obtain the time-domain signals after the separation of each normal mode, as shown in the figure below.

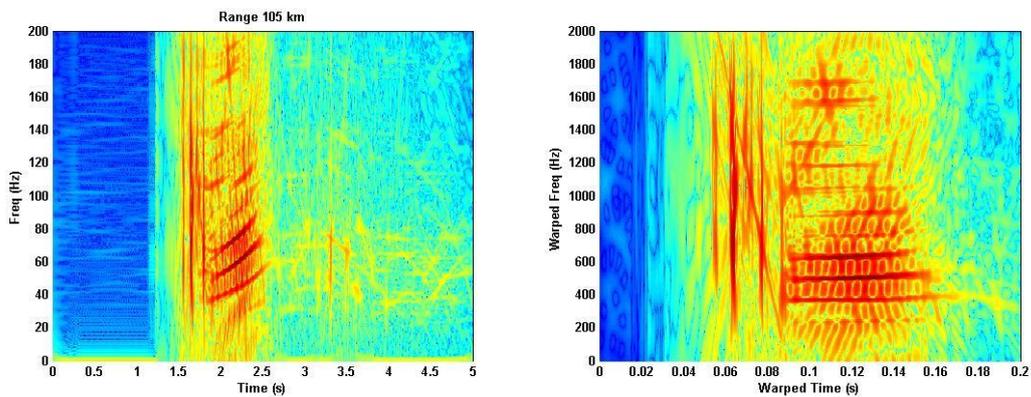

Figure 6 Warping transformation of a 105km signal

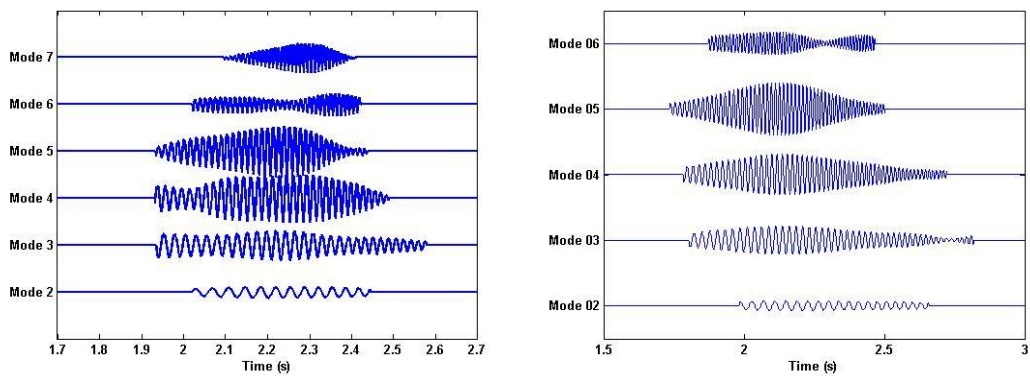

Figure 7 Time-domain signal waveform of each separated mode

After modal separation, the dispersion curves of each normal mode can be extracted using the time-domain waveform of each normal mode. By comparing the dispersion curves extracted after modal separation with the time-frequency analysis of the original signal, as shown in the figure below, it can be observed that the modal separation and dispersion curve extraction are relatively successful.

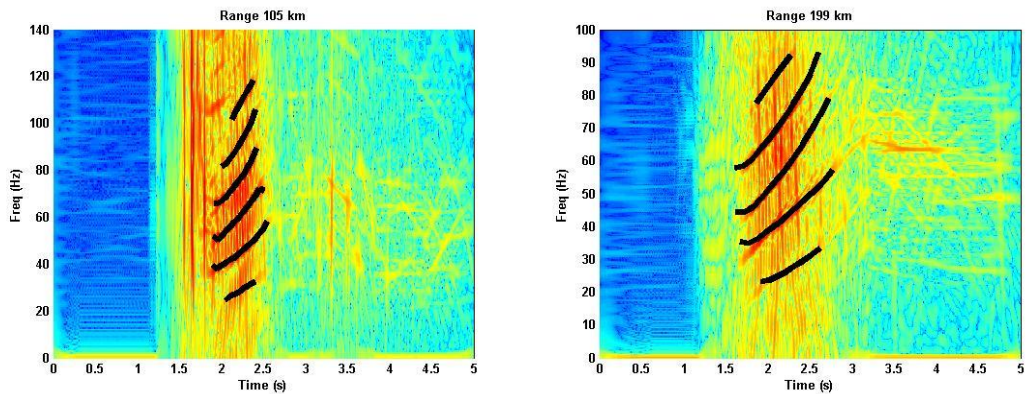

Figure 8 Frequency dispersion curves extracted from signals with propagation distances of 105km and 199km

By comparing the dispersion curve extracted from a sound source at a depth of 300m with the time-frequency analysis of a sound source at a depth of 100m, it can be observed that the dispersion structures of the two are consistent, but the amplitudes of the normal mode modes are not. From the comparison diagram, it can be seen that the depth of the sound source does not change the dispersion structure of the acoustic channel, but only changes the amplitudes of the excited normal modes with different numbers and frequencies. Therefore, the amplitude variation information of normal modes can be used to estimate the depth of the sound source.

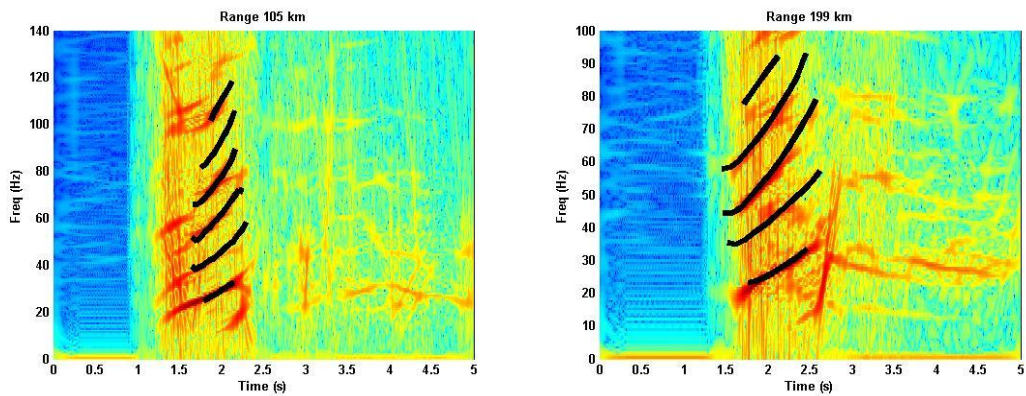

Figure 9 Frequency dispersion curves extracted from signals with propagation distances of 105km and 199km

# 4 Experimental results of sound source depth estimation

## A. Sound source depth estimation results based on normal mode amplitude information

The four acoustic signals with propagation range of 105km, 199km, 864km, and 1066km exhibit distinct multiple normal mode structures. Among them, at distances of 105km and 199km, there are two source depths: 100m and 300m, respectively. However, at distances of 864km and 1066km, there is only one source depth of 300m.

Below are the results of sound source depth estimation. The figure below presents the estimated sound source depths using acoustic signals with four different propagation distances when the sound source depth is 300m. As can be seen from the figure, the estimated depth is basically consistent with the depth of 300m, and the estimated depth is greater than the calibrated sound source depth of 300m.

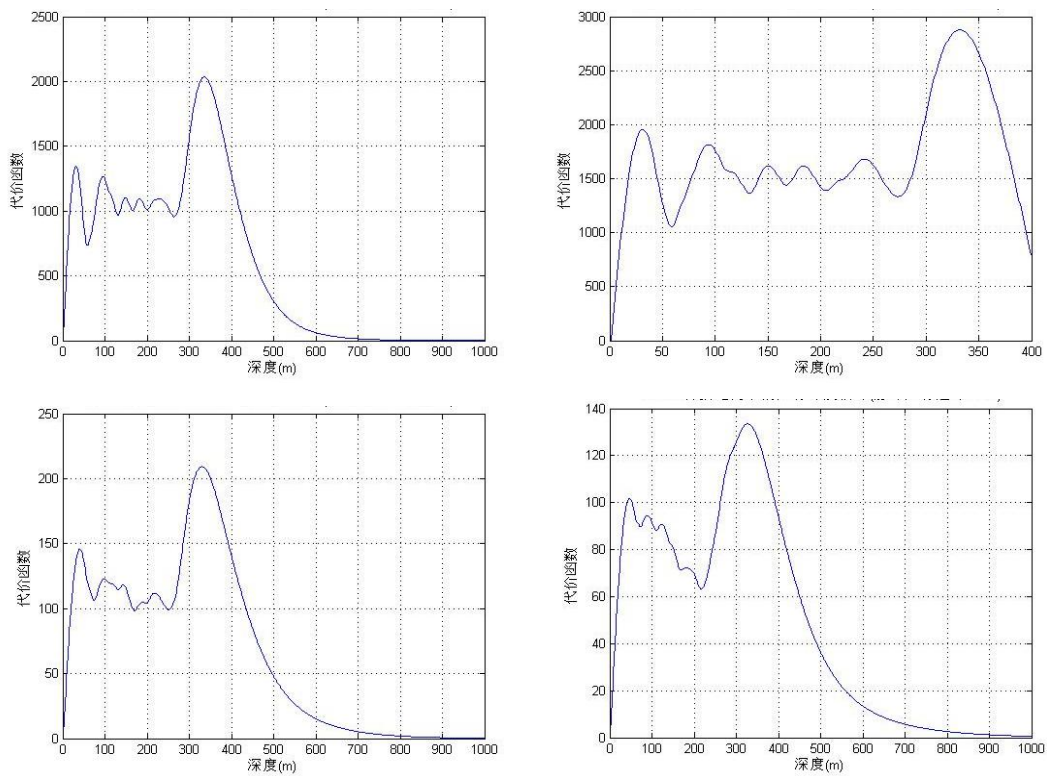

Figure 10 Depth estimation results based on normal modes (sound source depth is 300m)

The figure below presents the depth estimation results for a sound source depth of 100m. As can be seen from the figure, the estimation results exhibit multiple false peaks, yet the maximum value remains the actual depth. Compared to the case of a depth of 300m, multiple false peaks appear in the results for a depth of 100m. This is due to the eigenfunction exhibiting a pronounced periodic variation with depth at shallower depths, thereby forming multiple false peaks.

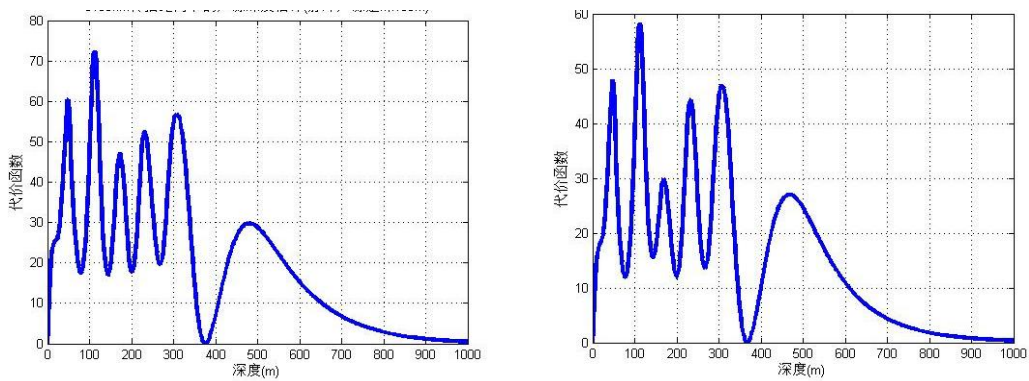

Figure 11 Depth estimation results based on normal modes (sound source depth is 100m)

## B. Sound source depth estimation results based on the upper frequency limit of normal modes

Based on the vertical distribution characteristics of eigenfunctions at different frequencies, when the sound source depth exceeds the vertical distribution range of the eigenfunction, the corresponding normal mode cannot be excited. Furthermore, the vertical distribution range of the normal mode eigenfunction gradually decreases as the frequency increases. Therefore, the sound source depth determines the upper frequency limit of the excited normal mode. In summary, the sound source depth can be inferred from the upper frequency limit of the normal mode. The figure shows the time-frequency analysis of received acoustic signals under different sound source depths. It can be seen from the figure that each normal mode has an upper frequency limit, and the upper frequency limit gradually decreases as the sound source depth increases. The simulation results verify the effectiveness of the method for estimating sound source depth based on the upper frequency limit.

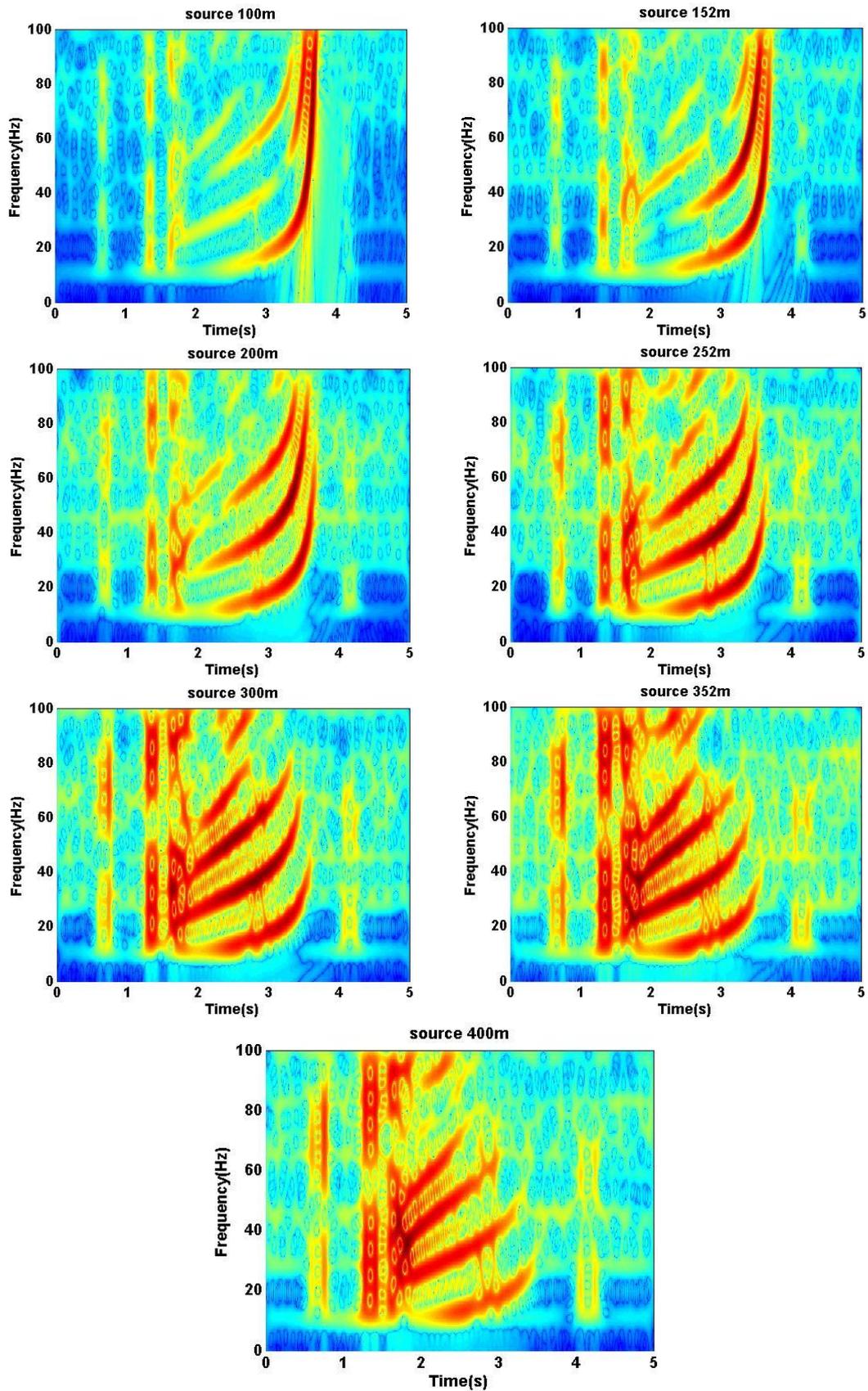

Figure 12 Time-frequency analysis of received acoustic signals under different source depths

By utilizing the normal mode sound field model and conducting calculations across multiple frequency points, one can obtain the curve depicting how the upper limit frequency of each normal mode varies with the depth of the sound source. As illustrated in the figure below, for the low-frequency range, the vertical distribution of the eigenfunctions for the first five normal modes within the 20Hz-100Hz band has been calculated. The maximum depth of the eigenfunction vertical distribution is obtained by limiting the amplitude to a minimum value. Since normal modes cannot be excited when the sound source depth exceeds the vertical distribution range of the eigenfunction, this result can be transformed into the upper limit frequency of each normal mode at a given sound source depth. In practical applications, normal mode sound field calculations are performed based on the measured sound velocity profile of the test sea area, obtaining the corresponding relationship curve between the upper limit frequency of normal modes and the sound source depth. Based on the upper limit frequency in the measured signal, the corresponding sound source depth can be quickly identified. This method is concise and efficient, and can be implemented using a single hydrophone.

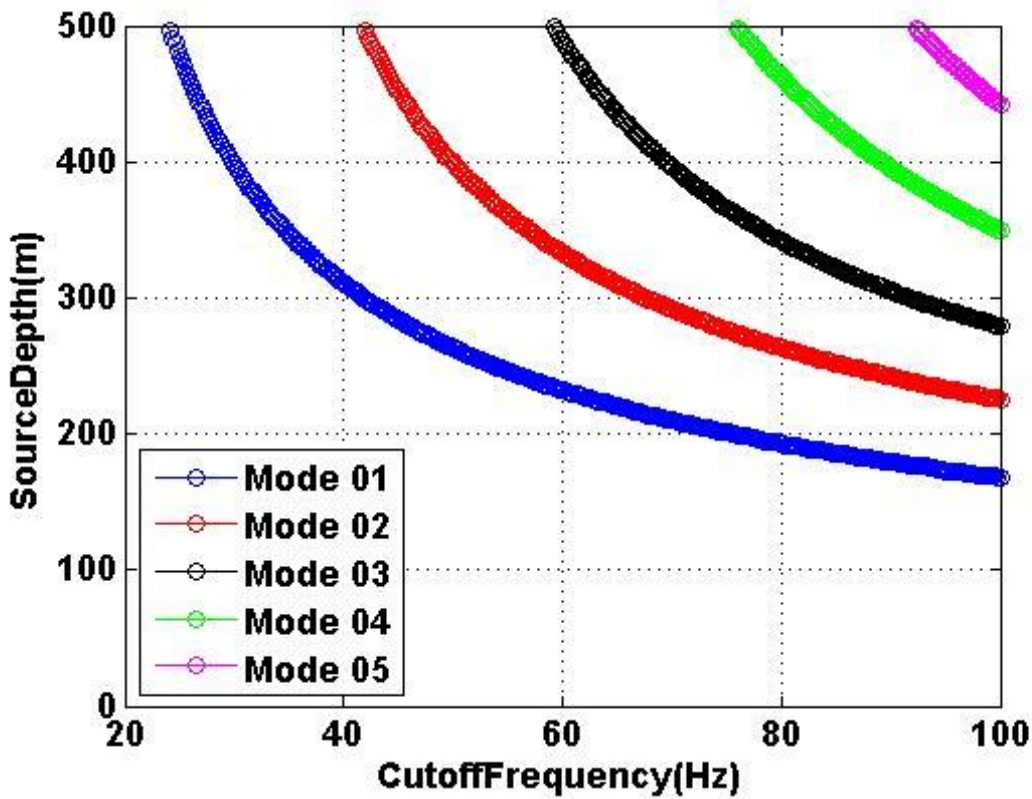

Figure 13 The relationship between source depth and modal upper limit frequency

Sound source depth estimation results. Taking simulated acoustic signals as an example, we conducted a demonstration of sound source depth estimation. Taking the first normal mode as an example, when the sound source depth is 200m, 300m, and 400m, the upper limits of modal frequencies are 80Hz, 40Hz, and 30Hz, respectively. By looking up the corresponding sound source depths in the graph, they are basically around 200m, 300m, and 400m. Taking the second normal mode as an example, when the sound source depth is 300m and 400m, the upper limits of modal frequencies are 70Hz and 50Hz, respectively. By looking up the corresponding sound source depths in the graph, they are basically around 300m and 400m. Taking the third normal mode as an example, when the sound source depth is 300m and 400m, the upper limits of modal frequencies are 90Hz and 70Hz, respectively. By looking up the corresponding sound source depths in the graph, they are basically around 300m and 400m. Therefore, this method is effective.

In addition, when the receiving depth is above the node of each normal mode, the receiving end cannot receive that normal mode, which can lead to errors in estimating the upper limit of modal frequency, such as the fourth and fifth normal modes. Therefore, in practical applications, it is necessary to consider the upper limits of frequencies of multiple modes and the impact of receiving depth to obtain an accurate estimate of the sound source depth.

## C. Sound source depth estimation results based on ray-based multi-path arrival

When both the source depth and the receiving depth are located at the surface layer of seawater, and the sound propagation path passes through deep-sea areas, the received acoustic signals can exhibit both the dispersion structure of multiple normal modes and the non-dispersion structure of multiple sound rays. The latter primarily involves sound rays arriving from deep-sea inversions or reflections from the seabed. The energy of sound rays reflected from the seabed is relatively weak and can be disregarded in long-distance scenarios. Therefore, in the case of single hydrophone reception, the time difference of multiple-path sound ray arrivals can be utilized for source depth estimation.

The figure presents the measured received acoustic signals at different distances. Only the acoustic signal with a propagation distance of 105km exhibits a distinct multi-path arrival structure, which is attributed to the obstruction effect of the seabed terrain. The acoustic sources are located at depths of 100m and 300m, respectively. As can be seen from the figure, different multi-path arrival structures are observed under varying acoustic source depths. By utilizing signal correlation, multiple pulse arrivals can be obtained. The figure below illustrates the multi-path arrival structure in the experimental data.

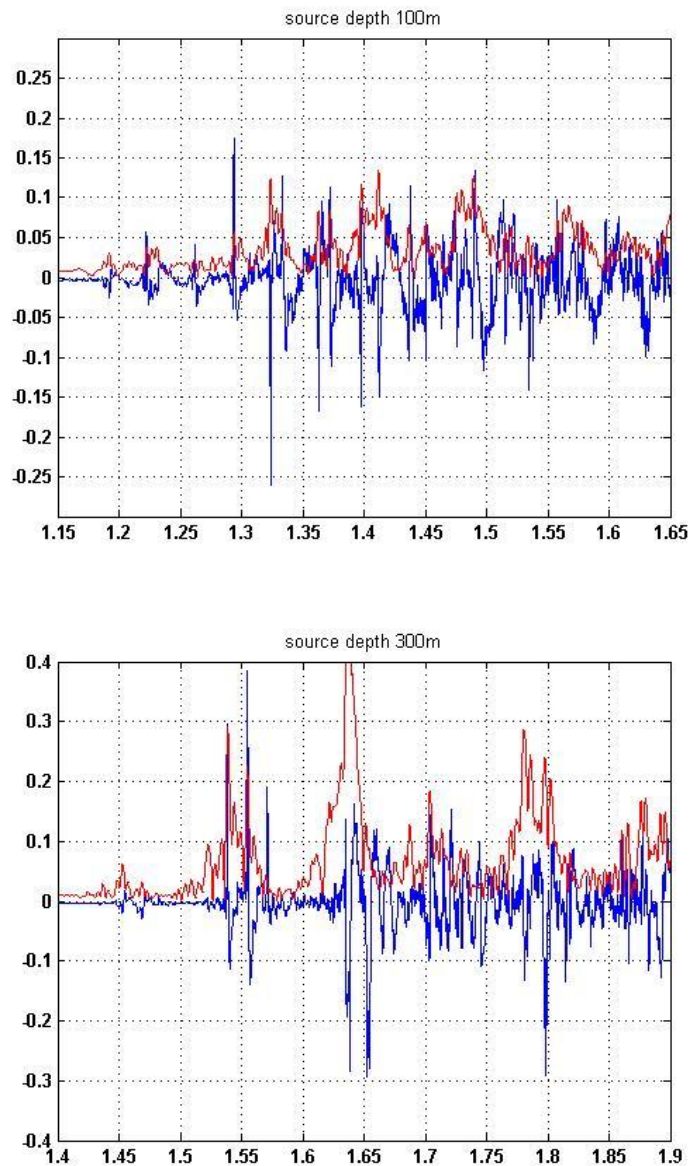

Figure 14 Signal time-domain waveform and signal correlation under conditions of sound source depths of 100m and 300m

Based on the sound speed profile and seabed topography measured in the experimental sea area, sound field simulation was conducted using a ray model to obtain the multi-path arrival structure and multi-path ray trajectories. The simulation results revealed that, in addition to complex multi-path arrival rays with small grazing angles at the surface, there were distinct rays that underwent three major depth reversals. Therefore, by utilizing the time difference of multi-path ray arrivals and given the known receiving depth, the source depth can be estimated by matching the multi-path time difference.

Based on the experimentally measured sound speed profile, ray trajectory and arrival structure simulations are conducted, considering both the ray trajectory in an ideal flat seabed terrain and the ray trajectory in an actual seabed terrain. For the ideal flat seabed condition, for an actual sound propagation distance of 105km, the deeply inverted rays need to undergo three deep inversions, with the earliest arriving rays being the sound source, the receiver, and the sea surface reflection.

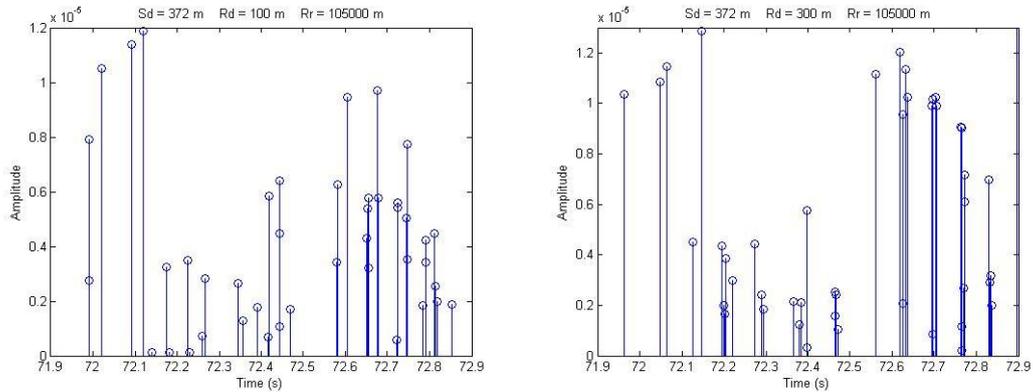

Figure 15 Sound ray arrival structures under ideal terrain conditions

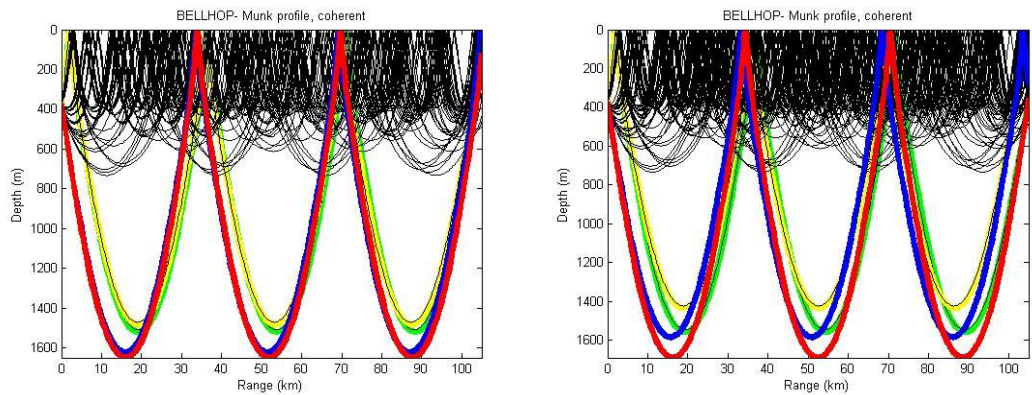

Figure 16 Sound ray trajectories under ideal terrain conditions

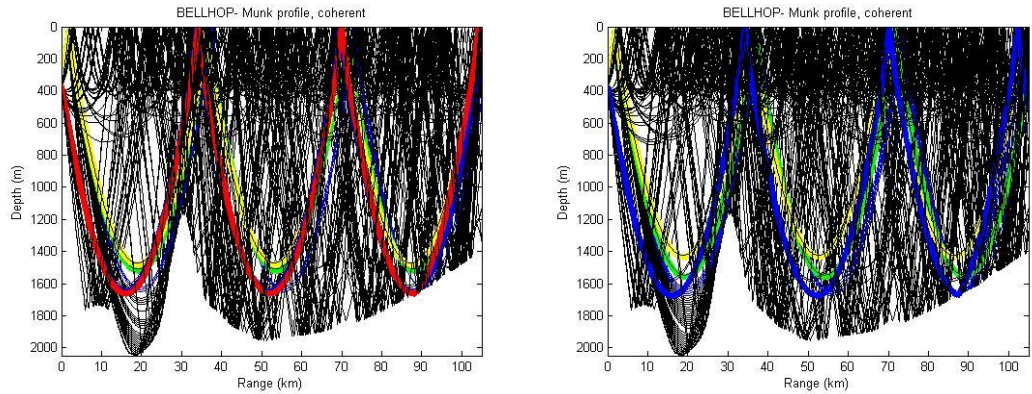

Figure 17 Sound ray trajectories under actual terrain

Comparing the multi-path arrival structure obtained from sound field simulation with the measured multi-path arrival structure, the results are shown in the figure below. It can be seen from the figure that the first four multi-path arrival structures in the measured signal are basically consistent with the simulation results.

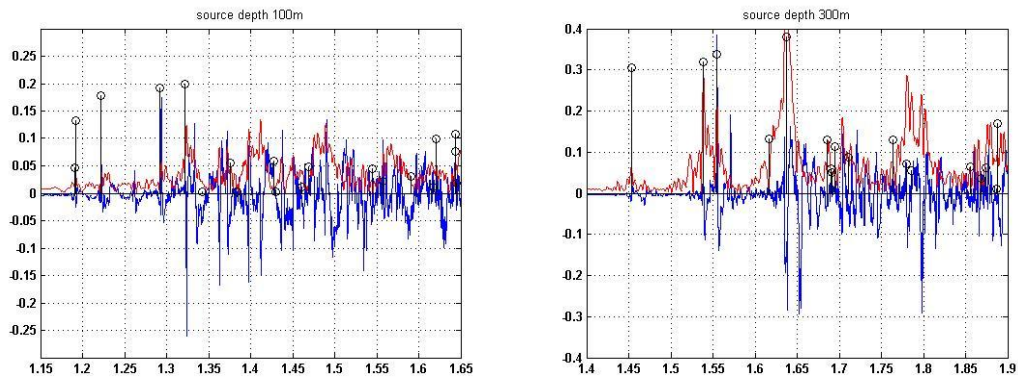

Figure 18 Comparison of wave arrival under actual terrain with sound ray arrival under ideal terrain

By simulating the multi-path arrival structure at different sound source depths, the time differences of the first four arriving sound rays are obtained. By matching these with the measured time differences of the four sound rays, the sound source depth estimation can be obtained.

## 5  Conclusion

In the case of an acoustic source within the surface layer of the deep Arctic sea, both normal modes and rays can be used to estimate the source depth. In the case of long distances, especially under sea ice coverage, the method using normal modes is applicable, as the ray arrival structure has become disordered. The ray method is limited by changes in seabed topography, while normal modes are not.

When both the sound source and the receiver are located at the surface layer of the Arctic deep sea, for low-frequency signals, a refractive-type normal mode waveguide can form at the surface layer, exhibiting a distinct modal dispersion structure. Therefore, the amplitude information of different normal modes at different frequencies can be obtained through modal separation. Simultaneously, within the surface layer waveguide, as the vertical coverage of the normal mode eigenfunctions decreases with increasing frequency, the amplitude of the normal modes varies significantly with frequency. This allows for the estimation of the sound source depth by matching the variation of normal mode amplitude with frequency, especially under conditions with multiple normal modes. In summary, based on the acoustic field characteristics of the refractive-type normal mode waveguide in the surface layer of the Arctic deep sea, single-hydrophone depth estimation of a low-frequency broadband pulse sound source can be achieved through modal separation, amplitude extraction, and matching of modal amplitude variation with frequency information using a single hydrophone.

When the sound source and the receiver are both located at the surface layer of the deep Arctic sea, within the refractive-like waveguide of the surface layer, the vertical coverage range of the eigenfunctions decreases with increasing frequency. When the depth of the sound source exceeds the vertical coverage range of the normal mode eigenfunctions, it is impossible to excite that particular normal mode. Therefore, under the condition of a fixed sound source depth, normal modes have an upper frequency limit. Thus, the estimation of the sound source depth can be obtained by matching the upper frequency limit of the normal modes. Of course, the prerequisite is that the depth of the receiver is less than the depth of the sound source; otherwise, the estimation result will be the depth of the receiver.

According to ray theory, when both the sound source and the receiver are located at the surface layer of the deep Arctic sea, the receiving end can collect the sound rays with strong energy that arrive earliest and have not been reflected by the seabed, including a cluster of four sound rays formed by the sound source, the receiver, and their reflections on the sea surface. The time difference between these four sound rays is determined by the depth of the sound source and the depth of the receiver. Therefore, given the known depth of the receiver, the depth of the sound source can be estimated by matching the time differences of the four sound rays. When the sound source is at the surface layer and the receiver is at a medium depth or near the seabed, the earliest arriving and strongest sound rays are the sound source-receiver and sound source-sea surface-receiver. Therefore, given the known depth of the receiver and the propagation distance, the depth of the sound source can be estimated based on the time difference.

If the signal acquisition device is a vertical array, vertical beamforming can be utilized to obtain the two arriving elevation angles. Through azimuth filtering and analysis of frequency domain fluctuations, the interference period between the sound source and the sea surface reflection can be obtained. Based on the elevation angle and interference period, the depth and distance of the sound source can also be estimated using ray theory. This method can also be applied to broadband continuous sound sources. At the same time, the array gain can be obtained using the array, which can expand the scope of the method. Furthermore, based on the vertical array, the modal energy time beam arrival structure can be obtained, which can be used to estimate the depth of the sound source. The vertical array can also be used to estimate the distance and depth of the sound source based on the matched field method. Therefore, two-dimensional sound source location estimation can be achieved using a vertical array.

If the signal acquisition device is a horizontal array, horizontal beamforming can be utilized to obtain the horizontal arrival beam angle. If the arriving signals include both normal modes and rays, two azimuth angles can be obtained. These two azimuth angles are related to the distance from the sound source and the horizontal azimuth, allowing for the estimation of the sound source distance and horizontal range. Additionally, based on the horizontal array, the modal energy time-beam arrival structure can be obtained, which can be used to estimate the depth of the sound source.

This method enables the estimation of the three-dimensional sound source location.

In summary, the sound field beneath the surface of the Arctic deep sea, originating from sound sources, exhibits both multipath arrival of sound rays and multimodal arrival of normal modes. This information can be fully leveraged to achieve superior results compared to previous methods. The aforementioned findings can serve as a guide for the design of sonar systems aimed at locating sound sources in the Arctic deep sea surface layer.

## Acknowledgments

The work presented in this article is supported by the National Key Research and Development Program, the National Natural Science Foundation of China, the Open Fund of National Key Laboratories, and the Basic Scientific Research Fund of the Third Institute of Oceanography.